# A Novel Approach for Invoice Management using Blockchain

Nikhil Sontakke[1], Shivansh Rastogi[1], Sejal Utekar[1], Shriraj Sonawane[1]
[1]Department of Computer Engineering, VIT Pune, India

**Abstract:-** Electronic invoicing is another area where blockchain technology is being used. Additionally, it has the power to alter how payments are made, invoices are issued, and transactions are validated. Using a blockchain-based invoicing system will enable smooth payments from a customer's digital wallet to a business's digital wallet. Transactions are simple to track and monitor, and the blockchain may be used to retrieve an exchange's full history. Sometimes shopkeepers create fake bills and submit them to the higher tax-paying authorities. To bring transparency to this billing system between customers, shopkeepers, and tax-paying authorities "billing system using blockchain" is to be implemented using the concept of Blockchain and make the billing system in our country work smoothly. Blockchain technology can revolutionize the invoicing and payment process by providing a secure, transparent and tamper-proof system. A blockchain-based billing system can facilitate smooth payments, allow for easy tracking and monitoring of transactions, and provide a tamper-proof history of all exchanges. The use of blockchain can prevent fraud and increase transparency among customers, shopkeepers, and tax-paying authorities. Furthermore, it can streamline the process by using digital wallets for both customers and businesses, reducing time and resources for traditional invoicing methods. Overall, blockchain technology can bring greater efficiency and trust to the billing system, benefiting all parties involved. It can prevent fraud, increase transparency and streamline the invoicing and payment process. This technology can create a more secure and efficient billing system ultimately benefiting all parties involved.

## I. INTRODUCTION

Almost all facets of modern life are made easier by technology advancements as all nations embrace the 4.0 industrial revolution. New developments brought forth by the technology's quick development offer enormous potential for market and commercial expansion [9]. Blockchain technology is a distributed, decentralized digital ledger that securely and openly records transactions. The technology that serves as the foundation for the virtual currency Bitcoin was first unveiled in 2008 by a person or group of persons named as Satoshi Nakamoto. [3]. Since then, the idea of blockchain has developed into one of the most anticipated technologies of the twenty-first century, with potential uses in a variety of sectors including finance, supply chain management, and healthcare.

Blockchain is an integrated innovation of already existing technologies, not a disruptive one. It combines smart contracts, distributed storage, consensus processes, and data encryption [6]. A chain of blocks, each containing a collection of transactions, is formed by the blockchain technology. Each block is linked to the one before it, creating a permanent record of every network transaction. The decentralized nature of the blockchain ensures that the ledger is tamper-proof, as any changes to a block would be immediately apparent to all parties on the network.

Low cost was cited by 52% of reviewers as the primary advantage of electronic invoices (e-invoices). Each paper invoice exchange costs e7, whereas e0.3 is charged for exchanges in electronic format, a 25-fold cost reduction. Additionally, each individual can only process 6,000 paper bills annually, whereas an individual can issue up to 90,000 invoices in electronic form. E-invoices will be successfully merged by the automatic digital system, which uses the e-invoice as input data [7]. Traditional payment methods are sometimes opaque, relying on paper, and leaving little or no audit trail. Debtors can easily avoid paying their debts by hiding behind bureaucracy or claiming that their claims have been delayed or lost. So using blockchain information can be accessible at each step of the transaction and transparency can be provided. The number of parties involved in the product transportation process and the number of handoffs that occur many times en route add to the complexity of global trade. Building end-to-end shipping visibility becomes quite difficult. Therefore, for the purposes of creating invoices, handling disputes, and settling payments, shippers and carriers are engaged to collect as much information as possible [5].

One of blockchain technology's most promising applications is the usage of bill management systems, which can increase efficiency and trust in the system by offering a safe and transparent means to record and verify transactions. In traditional bill management systems, it is easy for fraudulent activities to take place, such as shopkeepers creating fake bills and submitting them to tax-paying authorities. Blockchain allows for the recording of all transactions on a distributed ledger that is available to all parties. By doing this, all transactions are recorded in a visible and unchangeable manner, virtually eliminating the possibility of fraud. Furthermore, blockchain technology can also help to streamline the bill management process. Digital wallets can be used for both customers and businesses, allowing for easy and efficient payments. This can help to reduce the time and resources needed for traditional bill management methods, such as printing and mailing paper bills.The use of blockchain in bill management systems can bring greater transparency and trust between customers, shopkeepers, and tax-paying authorities.Additionally, it can aid in lowering the expenses and administrative strain connected to conventional bill management systems. The technology can help to prevent fraud and increase transparency, while also streamlining the bill management process.

Despite the many advantages of blockchain technology, there are also some challenges and limitations to its implementation. Scalability is a major issue because the network can currently only handle a certain amount of transactions. Another challenge is the lack of standardization and regulation, which can make it difficult for businesses to adopt the technology.

Finally, the advancement of blockchain technology has the potential to fundamentally change the way in which bill payment systems function. By providing a secure, transparent and tamper-proof system, blockchain technology can bring greater efficiency and trust to the bill management process. But there are also difficulties and constraints with its application that must be resolved. In order to understand the potential of blockchain technology for bill management systems, this research study will look at both its advantages and disadvantages. The paper will also discuss current developments and real-world examples of blockchain-based bill management systems, and provide insights into the future of this technology in this field.

## II. LITERATURE REVIEW

From the standpoint of OSCM, Rosanna Cole, Mark Stevenson, and James Aitken looked into blockchain technology. This article aims to lead to more research into blockchain technology from a management and operations perspective. Research agendas for the future will be developed based on identification of potential application areas. Different techniques, including strengthening product safety and security, enhancing quality management, lowering unauthorized counterfeiting, and enhancing sustainability, are possible to apply blockchain to OSCM operation. Additionally, it can decrease the need for middlemen and enhance supply chain interactions in a way that lowers costs. Inventory management and replenishment can also be improved [1]. Simanta Shekhar Sarmah provides background on Blockchain technology, its history, its architecture, its advantages, and its applications in a number of industries in this research-based paper where he discusses Blockchain technology, its history, its architecture, its workings, and its advantages and disadvantages. A major technological innovation in recent years has been blockchain technology. The blockchain has revolutionized the way businesses are conducted because it is a transparent system of money exchange. In the next five years, the blockchain market is predicted to be worth over 3 trillion dollars thanks to major investments from tech giants and corporations. The network consists of a digital ledger in a peer-to-peer network. It is gaining popularity because of its security and capability to solve digital identity issues.A simple introduction to blockchain technology is given in this application-based study by Arijit Chakrabarti and Ashesh Kumar Chaudhuri. Additionally, it explores how blockchain technology might greatly benefit customers and retailers by being applied to some business operations in the retail industry. The research highlights some of the challenges as well as the use of blockchain technology [3].

In a recently published application-based research article, Nam Ho Kim, Sun Moo Kang, and Choong Seon Hong proposed a mobile charger billing system for electric vehicles that makes use of Blockchain technology. Peer-to-peer online transactions are now more secure because of the application of this technology. Additionally, they examined the billing needs for mobile chargers and put up a lightweight solution to the problem of data size in the current Blockchain [4]. A blockchain-based e-invoice system for goods carriers is suggested in the paper "Blockchain Based e-Invoicing Platform for Global Trade" by Krishnasuri Narayanam, Seep Goel, et. al. It intends to increase the effectiveness and reduce the expenses of the e-invoicing process. The system promises to decrease disputes, expedite dispute settlement, and enable real-time auditing by using real-time shipment tracking data and pre-agreed service contract rates to generate invoices. The motivation for organizations to adopt such a system is to improve the overall efficiency and cost-effectiveness of global trade[5].

Blockchain technology may effectively address the problems of intermediaries' trust risk, reduce transaction costs, and boost synergy efficiency in a multi-agent context, according to Liu Xidong. The viability of using blockchain technology to create electronic invoices is discussed in this article, along with the general layout of the blockchain for electronic invoices. In the study, it is also recommended to build an alliance chain for electronic invoicing and employ intelligent contracts to put research ideas into practice for various alliance transactions [6].In this paper, Van-Cam NGUYEN, et. al suggest a method that uses Blockchain smart contracts to digitize invoices and calculate VAT automatically. The smart contract was created on the Remix IDE using the Solidity programming language and the Ethereum platform. According to empirical findings, the new model has cheap costs for digitizing invoices and figuring VAT. Our suggested strategy also lowers the danger of data loss attacks, which increases the credibility of the implementation of VAT payment (nonaffection from the third party) [7].

E-commerce system security, openness and trust, efficiency, and other specific challenges are now being addressed by the industry. These problems can be solved by implementing blockchain technology in the e-commerce industry. The potential uses of blockchain technology in the e-commerce industry were discussed in this paper. Examined in relation to blockchain applications and possibilities are several aspects of e-commerce, such as payment, security, supply chain, work automation utilizing smart contracts, and ethical standards for transparency in e-commerce transactions. [8]

This study examines how a VAT system can use blockchain technology, especially for electronic invoices (e-Invoice). This study used a qualitative methodology to explore blockchain technology models that could be used in a VAT system. The study's findings indicate that taxpayer data that doesn't need to be private can only be stored via blockchain technology. One example of data that is deemed secure if distributed across nodes in the blockchain technology network is the Tax Invoice Serial Number (TISN) [9].Chang, Yi-Wei, et. al. created an online marketplace powered by blockchain. They processed the money and secured the deposit using the self-enforcement of smart contracts. Each transaction is recorded in the decentralized ledger and blockchain-verified. Thus, trustless transactions are made possible. Without the involvement of reliable third parties, the smart contract can carry out reliable transactions, and blockchain transactions are traceable and irreversible. The blockchain stores all processes, including the introduction of the goods, the purchase, the delivery, and the payment. When a transaction dispute arises, it can be logged and utilized as electronic evidence in court [10].

This study looks at how smart contracts and blockchain technology can be used to efficiently bill for government services. The report also evaluates a number of government agencies and services to choose the appropriate blockchain type. Implementing blockchain and smart contracts reduces the problem of duplicate billing and payments, but it also has the potential to revolutionize the process by increasing the transparency of service billing and payment, which improves audit opportunities [11].In this paper, Guerar, Meriem, et.al. propose a model based on a public blockchain that permits fully open and group-restricted invoice auctioning. Furthermore, their strategy offers a reputation system based on the prior deeds of entities as documented on the open blockchain, enabling insurance providers to modify the cost of the insurance contracts they offer [12].

Distributed Ledger Invoice, a blockchain-based invoice discounting system, is introduced in this study, and a novel assessment method is suggested for assessing the present blockchain solutions for the invoice discounting scenario. Additionally, they go through two key challenges relating to interoperability and accessibility of information. Interoperability is crucial for blockchain's acceptance in interbanking operations because it is still a developing technology and multiple blockchain solutions may be employed in these activities. They also recommend a decoupling layer based on the Attribute-Based Access Control language to standardize access control to reserved information across several blockchains [13]. In this article, the block chain, which forms the basis of Bitcoin, is examined. BlockChain is a very appealing technology for resolving existing issues in the financial and non-financial industries because of its distributed ledger functionality and security.BlockChain-based business apps are quite popular, and as a result, many start-ups are developing them. The adoption will undoubtedly encounter the previously indicated severe headwinds. But even major financial organizations like Visa, Mastercard, Banks, and NASDAQ are making investments to investigate how to use current business models on BlockChain. In fact, several of them are looking for fresh company ideas in the blockchain industry. [14]

For SMEs to manage their liquidity concerns, factoring—where the invoice is cashed to prevent late payments from customers is a critical financial tool. Unfortunately, the fact that this business model relies on relationships with others and that the people involved in this case suffer from knowledge asymmetry make it unsafe. "Double funding" is one of the issues. which occurs when a SME draws money from several sources. They have proposed a system called DecReg that is built on blockchain technology in order to lessen this disparity and improve the scalability of this crucial instrument. We give performance analysis together with the protocols created for this framework .[1]

## III. SYSTEM DESIGN

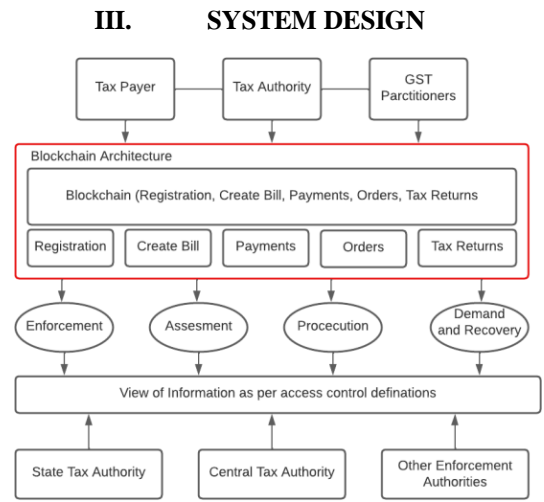

*Fig 1: Flowchart of Proposed Model*

The System is designed in such a way that the taxpayers can track their taxes and the government authorities monitoring taxation nationwide can track the shopkeepers' real payments with the help of all the billing data produced on the blockchain.

The Taxpayers and the authorities contribute directly to the blockchain. A single block consists of registration annual returns payments show cause notices and orders.

This data is then verified and used by/for law enforcement, assessment, deemed and recovery, and prosecution.

There are different views for the data stored in the blockchain: State Tax Authorities, Central Tax Authorities, and Other related authorities.

Views are basically used for giving access control to users based on specific data items.

## IV. METHODOLOGY

The project is built on a Web Application platform with HTML, CSS, and JavaScript, the three main languages used to build websites. Our website is programmed in JavaScript, structured in HTML, and styled using CSS. Bills are created via the UI, and bill data such as owner name, bill id, and so on are supplied as smart contract characteristics. A smart contract, which is a self-executing contract, directly incorporates the terms of the buyer-seller agreement into its lines of code. Smart contracts enable the implementation of reliable transactions and agreements between dispersed, anonymous parties without the need for a centralized authority, a legal framework, or an external enforcement mechanism. We are placing the data on the Ethereum Blockchain to assure its security. A peer-to-peer network for securely executing and verifying application code, or "smart contracts," is created by Ethereum, a decentralized blockchain platform. Without the aid of a reliable central authority, parties can conduct business with one another via smart contracts. We deployed the contracts on test networks before deploying them on the main network. We used Solidity to construct the contract. We used Remix IDE to create and deploy smart contracts, which are used to create a chain of transaction records and execute business logic in the blockchain system.

## V. IMPLEMENTATION

In implementation, the website frontend is created using Web3, a JavaScript library that allows users to interact with Ethereum blockchain. The first step is to fetch the contract that has been deployed on the blockchain. Once the contract is fetched, an instance of it is created. After creating the instance, the relevant information about the invoice, such as receipt number, total amount, seller identification, and buyer identification, is filled in as parameters. Once the parameters are filled, the event that has been created is described, and its emit function is called. By storing the variables on the blockchain with this emit function, they become unchangeable and tamper-proof. By enabling open access to all data stored on the blockchain, the system becomes transparent and trustworthy.

Additionally, the frontend can also include features such as user authentication, real-time updates, and notifications, to ensure smooth and efficient interactions with the blockchain contract. The frontend can also include a user-friendly interface that allows users to easily navigate and interact with the contract. Overall, the frontend serves as the bridge between the users and the blockchain contract, making it an important aspect of the overall implementation. The smart contract for the invoice generation system was created using Solidity, a programming language specifically designed for the Ethereum blockchain. RemixIDE, a web-based integrated development environment (IDE) that enables developers to build, test, and deploy smart contracts, was used to develop the contract.

For the implementation, the Ethereum Ropsten TestNet and Ganache were used as the blockchains. Ropsten TestNet is a test network for Ethereum, which allows developers to test their contracts without using real Ether. Ganache, on the other hand, is a localhost simulation of Ethereum, which allows developers to test their contracts in a local environment.

The website provides a great user interface for the shopkeeper or retailer and can provide with a lot of options like inventory management and order management apart from just invoice generation. This allows the shopkeeper or retailer to have complete control over their inventory and orders, which can help them keep track of the stock and avoid stockouts. The website also provides a lot of functionalities like real-time updates, notifications, and reports, which can help the shopkeeper or retailer to make better decisions.

Overall, the smart contract and the website together provide a robust and secure invoice management system that can help businesses to automate their invoicing process, improve their efficiency, and reduce the cost. Because the data is immutable and tamper-proof thanks to the usage of blockchain technology, the system is more transparent and trustworthy. Additionally, the user-friendly interface of the website makes it easy for users to interact with the contract, making it accessible for businesses of all sizes.

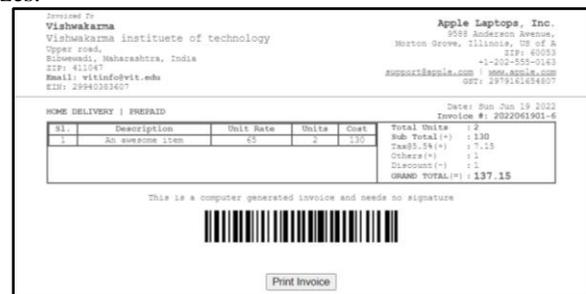

*Fig 2: Generated Invoice*

Pseudo Code for smart contract in Solidity for creating and paying the bill:
1. Initialize an empty list of bills
2. Initialize a bill counter to 0
3. Create a function called "create bill"
   a. Input: payee's address, bill amount, and memo
   b. Increment the bill counter
   c. Create a new bill with payee's address, bill amount, memo, and "unpaid" status
   d. Add the new bill to the list of bills
   e. Emit an event "bill created" with bill ID, payee's address, bill amount and memo

4. Create a function called "pay bill"
 a. Input: bill ID
  b. Fetch bills from list of bills, using bill ID
  c. Check if the bill is unpaid
 i. If it is unpaid, check if the msg.value is equal to the bill amount
   1. If true, transfer the bill amount to the payee's address
   2. Mark the bill as paid
   3. Emit an event "bill paid" with bill ID

We also deployed the contracts on Ganache which is a LocalHost simulation of Ethereum.

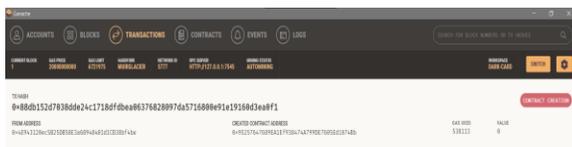

*Fig 3: Contract deployed on Ganache*

## VI. CONCLUSION AND FUTURE SCOPE

In this article, we covered blockchain technology and how it may be used as a billing system in the retail industry. By enhancing the transparency of products and overall information of bill generation the retail industry can be benefitted by blockchain technology.

This system will minimize present tax evasion, and the government will be able to track it. In addition to that, more openness will be offered to customers so that they may learn whether the tax they pay for a product is truly paid to the government**.**